\documentstyle[aps,psfig,epsf,preprint]{revtex}

\begin{document}

\draft

\title{$CP$-violating asymmetry in $\Lambda\to p\pi$ in the Skyrme model}
\author{Ji-Hao Jiang \footnote{jjh@mail.ustc.edu.cn} and Mu-Lin Yan
\footnote{mlyan@ustc.edu.cn}}
 \address{Interdisciplinary Center for Theoretical Study\\
University of Science and Technology
 of China, Hefei, Anhui 230026 China}

\begin{abstract}
We study the $CP$-violating asymmetry in nonleptonic decay $\Lambda\to
p\pi$. By employing the Skyrme model to calculate this decay amplitude
contributed by the gluonic diploe operator, we find a possible large
$CP$-violating asymmetry could be expected, which is consistent with the
previous study.
\end{abstract}

\maketitle

\section*{}
Hyperon decays\cite{{X.G. He PRD61},{Jusak Tandean},{hyperon1}} play an
important role in both studying the $CP$-violating phenomena in particle
physics and searching for new physics beyond the standard model. The
$CP$-violating asymmetry in decay $\Lambda \to p\pi$ has been investigated
in Ref.\cite{{X.G. He PRD61},{Jusak Tandean}}, and the asymmetries
$A(\Lambda^0_-)$ in $\Lambda \to p\pi$ have a simply form when the small
$\Delta I=3/2$ amplitude is neglected \cite{J. Donoghue}
\begin{eqnarray}
A(\Lambda_-^0)=\frac{\alpha+\bar{\alpha}} {\alpha-\bar{\alpha}}\approx -{\rm
tan}(\delta_{p}-\delta_s){\rm sin}(\phi_p-\phi_s), \label {A-define}
\end{eqnarray}
where $\alpha$ is hyperon decays parameter of $\Lambda^0_- \to
p\pi^-$ and $\bar{\alpha}$ corresponds to $\bar{\Lambda^0}_- \to
\bar{p}\pi^+$, $\delta_s=6^o$ and $\delta_p=-1.1^o$ are the final
state $\pi N$ strong interaction phases for $S$ and $P$ wave
amplitudes with $\Delta I=1/2$ respectively\cite{L. D. Roper}, and
$\phi_{s,p}$ are the corresponding $CP-$violating weak phases. The
standard model prediction for $A(\Lambda^0_-)$ is around $3\times
10^{-5}$, and the recent study in Ref.\cite{Jusak Tandean} shows
\begin{eqnarray}
-3\times 10^{-5}\leq A(\Lambda_-^0) \leq 4\times 10^{-5}.
\end{eqnarray}
A model independent study of new $CP-$violating interactions has shown that
$A(\Lambda_-^0)$ could be larger than that predicted within the standard
model\cite{X.-G. He 1995}. An example of an operator is precisely the gluonic
dipole operator\cite{{X.G. He PRD61},{X.-G. He 1995}}, in which
$A(\Lambda^0_-)$ would be enhanced and could reach around
$O(10^{-3})$\cite{X.G. He PRD61}.

The short-distance effective Hamiltonian for the gluonic dipole operator
is\cite{Buras:1999da}
\begin{eqnarray}
{\cal{H}}_{eff}=C_g^+Q_g^++C_g^-Q_g^-+{\rm h.c.},
\end{eqnarray}
where
\begin{eqnarray}
Q_g^{\pm}&=&\frac{g} {16\pi^2} ({\bar{s}_{\rm
L}\sigma^{\mu\nu}t^aG^a_{\mu\nu}d_{\rm R}\pm \bar{s}_{\rm
R}\sigma^{\mu\nu}t^aG^a_{\mu\nu}d_{\rm L}}),
\end{eqnarray}
and in supersymmetric model the dominant contribution to Wilson
coefficients generated by gluino exchange
diagrams\cite{{Buras:1999da},{Gabbiani:1996hi}} are given by
\begin{eqnarray}
C_g^{\pm}=\frac{\pi \alpha_s(m_{\tilde{g}})} {m_{\tilde{g}}}
[(\delta_{LR}^d)_{21}\pm (\delta_{LR}^d)_{12}^*]G_0(x).
\end{eqnarray}
Here $(\delta_{\rm LR}^d)_{ij}=(M^2_d)_{i_Lj_R}/m^2_{\tilde{q}}$ denote the
off-diagonal entries of the (down-type) squark mass matrix in super-CKM
basis\cite{{Gabbiani:1996hi},{L.J. Hall}} and
$x=m^2_{\tilde{g}}/m^2_{\tilde{q}}$ the ratio of gluino and squark mass
squared. The loop function $G_0(x)$ is\cite{G. Colangelo}
\begin{eqnarray}
G_0(x)=x\frac{22-20x-2x^2+(16x-x^2+9){\rm log}x} {3(x-1)^4}.
\end{eqnarray}
Note that $G_0(1)=-5/18$ and the function does not depend strongly on $x$.

The $CP$-violating asymmetry $A(\Lambda^0_-)$ generated by the
gluonic dipole operator has been calculated in Ref.\cite{X.G. He
PRD61}, and a large enhancement of the asymmetry could be expected
in the supersymmetric extensions of the standard model. It is
known that models are necessary for evaluating the hadronic matrix
elements $<p\pi|Q_g^\pm|\Lambda>$ to predict the asymmetry
$A(\Lambda_-^0)$. The MIT bag model was used for this task in
Ref.\cite{X.G. He PRD61}. The purpose of this letter is to
recalculate the hadronic matrix elements
$<p\pi|Q^{\pm}_g|\Lambda>$ by using the Skyrme model, then to
evaluate the asymmetry $A(\Lambda_-^0)$, in order somewhat to
complement the work by the authors of Ref.\cite{X.G. He PRD61}.

Generally the $SU(3)$ extended Skyrme Lagrangian is\cite{{skyrme},{H. Weigel}}:
\begin{eqnarray}
{\cal L}&=&\frac{1}{16}f_\pi^2{\rm Tr}(\partial_\mu U\partial^\mu U) +
\frac{1}{32e^2}{\rm Tr}([\partial_\mu UU^\dagger,\partial_\nu
UU^\dagger]^2) \nonumber \\
&+&\frac{f_\pi^2}{16}m^2{\rm Tr}(U+U^\dagger-2)
-\frac{f_\pi^2}{8\sqrt{3}}\Delta m^2{\rm Tr}(\lambda_8(U+U^\dagger))
\label{skyrme}
\end{eqnarray}
where $U=\exp(i2\lambda_i \phi_i/F_\pi)$, $\phi_i$ are octet pseudoscalar
fields;
and $m^2=(3m_\pi^2+4m_K^2+m_\eta^2)/8,\;  \Delta
m^2=m^2_K-m^2_\pi$. The third and fourth term in eq.(\ref{skyrme})
describe the chiral symmetry breaking and the SU(3) flavor
symmetry breaking respectively. The model has only one free
parameter, namely, the Skyrme parameter $e$, which could be fixed
phenomenologically. In the Skyrme model, baryon fields emerge as
topological soliton in the pseudoscalar meson fields
theory\cite{{skyrme},{H. Weigel}}, and the space-time-dependent
matrix field $U_B(\vec r, t)\in SU(3)$ takes the form
\begin{eqnarray}
U_B(\vec r,t)=A(t)U_0(\vec r)A^\dag(t)
\end{eqnarray}
where $U_0(\vec r)$ is the SU(3) matrix, and $A$ is arbitrary time-dependent
$SU(3)$ matrix. Using hedgehog ansatz\cite{{skyrme},{H. Weigel}}, we have
\begin{eqnarray}
U_0=\left[ \begin{array}{cc} e^{iF(r)\vec{r}\cdot\tau} & 0 \\
0 & 1 \end{array} \right ]=\left[\begin{array}{cc} \cos(F) +i
\vec{r}\cdot\tau\sin(F) & 0\\
0 & 1 \end{array} \right ],
\end{eqnarray}
where the profile function $F(r)$ is chiral angle that
parameterizes the solution, which satisfies the following
equation of motion
\begin{equation}
(\frac{x^2}{4}+2S^2)F''+\frac{xF'}{2}+2SC(F'^2-\frac{1}{4}-\frac{S^2}{x^2})-\frac{1}{4e^2f_\pi^2}(m^2-\frac{2}{3}
\Delta m^2)x^2S=0, \label{eq.F}
\end{equation}
with the boundary conditions $F(0)=\pi, \;F(\infty)=0$, where
$C=\cos(F)$, $S=\sin(F)$ and $x=f_\pi r$. The last term of the
equation of motion (\ref{eq.F}) responds to the SU(3) symmetry
breaking due to $m_K \neq m_\pi$ (or $m_s\neq (m_u,\;m_d)$) at the
classical soliton solution level. In this letter, only this sort
of SU(3) symmetry breaking effects is taken into account, and
one's at the soliton's semiclassical quantization level will be
ignored (see below).

Considering pion fluctuations on a classical solution, we define the pion as
chiral perturbations about $U_B$\cite{H.J. Schnitzer}
\begin{eqnarray}
U=U_\pi U_B(\vec{r},t)U_\pi
\end{eqnarray}
where
\begin{eqnarray}
U_\pi=\exp(i\lambda_i \cdot \pi_i /f_\pi), \label{Upi}
\end{eqnarray}
and $\pi_i, i=1,2,3$, is the pion field. Thus by expanding
Eq.(\ref{Upi}), one can calculate the amplitude of hyperon
non-leptonic decay amplitude. In this letter, we will use
``semiclassical" approximation associated with the fact that
solitons are slowly rotating. Within this approach, we do not need
consider all time derivative times.

The realization of $Q_g^{\pm}$ in terms of meson fields, to the
leading order in $1/N_c$ and in the derivative (or $p-$)expansion,
can be written as\cite{G.D'Ambrosio}
\begin{eqnarray}
Q_g^{\pm}=\frac{11} {256 \pi^2} \frac{f_{\pi}^2m_K^2} {m_s+m_d} \times
\frac{1}{2}{\rm Tr}[(\lambda_6-i\lambda_7)(U\partial_\mu U^\dag
\partial^\mu U\pm \partial_\mu U^\dag \partial^\mu U U^\dag)]. \label{QQQ}
\end{eqnarray}
Since the transfer momentum $p$ for $\Lambda \rightarrow p\pi$ is
about $m_\Lambda-m_p-m_\pi \sim 0.035GeV$, which is much smaller
than the chiral symmetry spontaneously breaking scale
$\Lambda_\chi \simeq 2\pi f_\pi \simeq 1.2 GeV$, the ${\cal
O}(p^2)$-contributions of $Q_g^\pm$ to the decay
$\Lambda\rightarrow p\pi$ are dominant and the corrections of
${\cal O}(p^4)$'s should be relatively small and ignorable at this
stage. In other words, as an approximation expression of
$Q_g^\pm$, Eq.(\ref{QQQ}) is good enough for our purpose. Note
that the overall factor of Eq.(\ref{QQQ}), which cannot be fixed
model-independently, is obtained by using chiral quark
model\cite{Bertolini:1994qk}.

Since there is only one $\pi$ meson in the matrix element $<p\pi^-|{\cal
H}_{eff}|\Lambda^0>$, we only consider the term with one $\pi$ term in
$Q_g^\pm$.
\begin{eqnarray}
\lefteqn{Q_g^\pm=-\frac{i\pi_j}{2f_\pi}\frac{11}{256\pi^2}\frac{f_\pi^2m_K^2}{m_s+m_d}} \nonumber \\
& & \times {\rm Tr}\{ (\lambda_j\lambda_{6-i7} +\lambda_{6-i7}\lambda_j) (\partial_iU_B \partial^i U_B^\dagger U_B\mp U_B\partial_i U_B^\dagger \partial^i U_B)\} \nonumber \\
& &-\frac{i \partial_i \pi_j}{2 f_\pi}\frac{11}{256\pi^2}\frac{f_\pi^2m_k^2}{m_s+m_d}\nonumber \\
& &\times {\rm Tr} \{\lambda_{6-i7}(-\lambda_j \partial^i U_B^\dagger- \partial^i U_B \lambda_j -\partial^i U_B^\dagger U_B \lambda_j U_B^\dagger +U_B^\dagger \lambda_j \partial^i U_B U_B^\dagger)\nonumber \\
& &\ \ \pm \lambda_{6-i7}(\lambda_j \partial^i U_B + \partial^i U_B
\lambda_j +U_B \lambda_j U_B^\dagger \partial^i U_B -U_B \partial^i
U_B^\dagger \lambda_j U_B)\}
\end{eqnarray}
where notations $\lambda_{6-i7}\equiv\lambda_6-i\lambda_7$ etc.
have been used. Then it is straightforward to get the
corresponding $S$-wave and $P$-wave amplitudes as follows:
\begin{eqnarray}
s = -\frac {11} {256\pi^2} {m_{\tilde g}} \frac {f_{\pi}^2m_K^2}
         {m_s+m_d}& &C^{-*}_g \frac{2\sqrt{2}} {f_{\pi}}D_{4+i5,8}I, \label{s}\\
p =\frac {11} {256\pi^2} {m_{\tilde g}} \frac {f_{\pi}^2m_K^2}
        {m_s+m_d} |\vec{q}| & &\frac{2\sqrt{2}} {f_\pi}
     C^{+*}_g\{D_{4+i5,3}I_1-(D_{4+i5,3}(D_{3,8}-\frac{1}{\sqrt{3}}D_{8,8})+
     \nonumber \\
    & &  D_{4+i5,8}(D_{3,3}-\frac{1}{\sqrt{3}}D_{8,3}))I_2
     +D_{4+i5,j}(D_{3,n}-\frac{1}{\sqrt{3}}D_{8,n})i\epsilon_{3jn}I_3\}, \label{p}
\end{eqnarray}
where
\begin{eqnarray}
D_{ab}& = &\frac{1}{2}Tr[\lambda_a A \lambda_b A^\dag], \nonumber \\
D_{4+i5,b}&=&D_{4,b}+iD_{5,b}, \nonumber \\
I& = &\int d^3x \frac{C}{\sqrt{3}}(\frac{2S^2}{r^2}+{F'}^2), \nonumber \\
I_1& = &\int
d^3x(\frac{1}{3}(\frac{2SC}{r}+CF')+\frac{2C+1}{9}(\frac{2SC}{r}+F')),\nonumber\\
I_2& = &\int d^3x\frac{C-1}{3\sqrt{3}}(\frac{2SC}{r}+F'),\nonumber \\
I_3& = &\int d^3x\frac{2S^3}{3r}.
\end{eqnarray}
and $j,n=1,2,3$. Some hadronic matrix elements of D-function that will be
used in below read
\begin{eqnarray}
& &\left<p \downarrow|D_{4+i5,8}|\Lambda \downarrow
\right>=-\frac{2}{5},\;\;\; \left<p \downarrow|D_{4+i5,3}|\Lambda
\downarrow\right>=-\frac{2\sqrt{6}}{15}
\\
& & \left<\right.p \downarrow|D_{4+i5,3}(D_{3,8}-\frac{1}{\sqrt{3}}D_{8,8})|\Lambda \downarrow\left.\right>=-\frac{2}{75},\\
& & \left<\right.p \downarrow|D_{4+i5,8}(D_{3,3}-\frac{1}{\sqrt{3}}D_{8,3})|\Lambda \downarrow\left.\right>=-\frac{3\sqrt{2}}{100} \\
& &i \epsilon_{3jn}\left<\right.p
\downarrow|D_{4+i5,j}(D_{3,n}-\frac{1}{\sqrt{3}}D_{8,n})|\Lambda
\downarrow\left.\right>=-\frac{\sqrt{6}}{50}.
\end{eqnarray}
In the derivations of the above quantities, we have taken
SU(3)-$D_{\mu,\nu}$-functions to be the baryon wave functions.
This implies that the SU(3)-symmetry breaking effects at the
soliton's semiclassical quantization level are not taken into
account at this stage. To a further estimation to corrections due
to SU(3)-asymmetric wave functions of baryons, one could employ
the Yabu-Ando's method\cite{Yabu} to do so, which however exceeds
the contents of this present letter. In this letter we only take
the SU(3)-symmetry breaking effects due to classical soliton
equation eq.(\ref{eq.F}) into account.

Since the Wilson coefficients $C_g^\pm$ could be complex, it is
easy to extract the imaginary part amplitudes of the decay from
Eqs.(\ref{s}) and (\ref{p}). By taking the real part of the
amplitudes from experimental data, $s_{\rm expt}=3.3\times
10^{-7}, p_{\rm expt}=1.2\times 10^{-7}$\cite{PDG2002}, and using
$m_s+m_d=0.110GeV, m_{\tilde{g}}\sim 500GeV\cite{X.G. He PRD61,M.
Ciuchini JHEP10}, m_K=0.495GeV$, we get the phase of $S$-wave and
$P$-wave as
\begin{eqnarray}
\phi_s & = &(3.3\times 10^{-7})^{-1}\cdot(-6.68\times 10^{-6})
\times{\rm Im}[(\delta_{RL}^d)_{12}-(\delta_{LR}^d)_{12}], \nonumber \\
\phi_p & = &(1.2\times 10^{-7})^{-1}\cdot (3.82 \times 10^{-6})\times{\rm
Im}[(\delta_{RL}^d)_{12}+(\delta_{LR}^d)_{12}].
\end{eqnarray}
Thus we find
\begin{eqnarray}
A(\Lambda_-^0)& = &6.2{\rm Im}(\delta_{RL}^d)_{12}+1.4{\rm
       Im}(\delta_{LR}^d)_{12}. \label{our.result}
\end{eqnarray}
The result in Ref.\cite{X.G. He PRD61} is
\begin{eqnarray}
A(\Lambda^0_-)=(2.0B_p+1.7B_s){\rm Im}(\delta_{RL}^d)_{12}
                +(2.0B_p-1.7B_s){\rm Im}(\delta_{LR}^d)_{12}.\label{re1.result}
\end{eqnarray}
It is shown in Ref.\cite{X.G. He PRD61} that $B_p$ and $B_s$
quantify the uncertainty in the matrix elements
$<p\pi|Q_g^\pm|\Lambda>$ with $0.5<B_s<2.0, 0.7B_s<B_p<1.3B_s$,
and $A(\Lambda^0_-)$ could be range of $O(10^{-3})$ without
conflict with other constraints. Consider the range of $B_s$ and
$B_p$, it is not difficult to see that our result
[Eq.(\ref{our.result})] is consistent with theirs
[Eq.(\ref{re1.result})]. Using the similar analysis, we can
therefore expect that $A(\Lambda_-^0)$ could also be $O(10^{-3})$
in the present letter.

In conclusion, we have restudied the supersymmetric contribution to the
$CP$-violating asymmetry in hyperon non-leptonic decay $\Lambda\to p\pi$.
Different from Ref.\cite{X.G. He PRD61}, in order to predict the asymmetry
$A(\Lambda_-^0)$, we use the Skyrme model to calculate the hadronic matrix
elements $<p\pi|Q_g^\pm|\Lambda>$. It is found that the result is
consistent with that given in Ref.\cite{X.G. He PRD61}, and
$A(\Lambda^0_-)$ could be range of $O(10^{-3})$.

\section*{acknowledgements}
We thank Dao-Neng Gao for useful discussions. This work is partially
supported by NSF of China 90103002.

\end{document}